\documentclass[prd, twocolumn]{revtex4}

\usepackage{amssymb, amsmath} 
\usepackage{stmaryrd}

\usepackage{bbm} 

\usepackage[english]{babel}
\usepackage{slashed}
\usepackage{cancel}

\usepackage{slashbox}

\usepackage{graphicx}
\usepackage{color}

\usepackage{datetime}
\settimeformat{ampmtime}

\usepackage{float}

\usepackage[bookmarks,breaklinks,plainpages=false,pdfpagelabels]{hyperref} 
	\hypersetup{linkcolor=red,citecolor=blue,filecolor=dullmagenta,urlcolor=darkblue} 

\definecolor{grey}{rgb}{0.4,0.4,0.4}
\definecolor{lightgrey}{rgb}{0.6,0.6,0.6}
\definecolor{dullmagenta}{rgb}{0.4,0,0.4}
\definecolor{darkblue}{rgb}{0,0,0.4}
\definecolor{orange}{rgb}{1,0.5,0}
\definecolor{lightbrown}{rgb}{0.75,0.5,0.25}
\definecolor{tan}{cmyk}{0.14,0.42,0.56,0}
\definecolor{djunglegreen}{cmyk}{0.99,0,0.52,0}
\definecolor{lightgreen}{rgb}{0,1,0}
\definecolor{olivegreen}{cmyk}{0.64,0,0.95,0.40}
\definecolor{midgreen}{rgb}{0.0,0.675,0.0}

\newcommand{\qq}{\qquad}

\newcommand{\vs}{\vspace}

\renewcommand{\.}{\hspace{0.5mm}}

\newcommand{\Krm}{\ensuremath{\mathrm{K}}}

\newcommand{\Wrm}{\ensuremath{\mathrm{W}}}

\newcommand{\grm}{\ensuremath{\mathrm{g}}}

\newcommand{\irm}{\ensuremath{\mathrm{i}}}

\newcommand{\krm}{\ensuremath{\mathrm{k}}}

\newcommand{\mrm}{\ensuremath{\mathrm{m}}}

\newcommand{\srm}{\ensuremath{\mathrm{s}}}

\newcommand{\Hcal}{\ensuremath{\mathcal{H}}}

\newcommand{\Kcal}{\ensuremath{\mathcal{K}}}

\newcommand{\Ocal}{\ensuremath{\mathcal{O}}}

\newcommand{\Ucal}{\ensuremath{\mathcal{U}}}

\newcommand{\jbm}{\ensuremath{\bm{j}}}

\renewcommand{\d}{\ensuremath{\mathrm{d}}}
\newcommand{\ee}{\ensuremath{\mathrm{e}}}

\newcommand{\defas}{\mathrel{\mathop :}=} 
\newcommand{\hph}[1]{\hphantom{#1\;\,}}

\newcommand{\eg}{e.g.}
\newcommand{\ie}{i.e.}

\newcommand{\rhs}{r.h.s.}

\newcommand{\cf}{cf.}

\makeatletter
 \def\ifundefined#1{\expandafter\ifx\csname#1\endcsname\relax}
 \ifundefined{default@color}
  \let\default@color=\current@color
 \fi
\makeatother

\newcommand{\beq}{\begin{equation}}
\newcommand{\eeq}{\end{equation}}
\newcommand{\bea}{\begin{eqnarray}}
\newcommand{\beas}{\begin{eqnarray*}}

\newcommand{\eea}{\end{eqnarray}}
\newcommand{\eeas}{\end{eqnarray*}}

\usepackage{simplewick}

\begin{document}

\title{Bose-Einstein Condensates with Derivative and Long-Range Interactions\\as Set-Ups for Analog Black Holes}

\author{Florian K{\"u}hnel}
	\email{florian.kuhnel@fysik.su.se}
	\affiliation{The Oskar Klein Centre for Cosmoparticle Physics,
			Department of Physics,
			Stockholm University,
			AlbaNova,
			106\.91 Stockholm,
			Sweden}

\date{\formatdate{\day}{\month}{\year}, \currenttime}

\begin{abstract}
General types of Bose-Einstein condensates are considered. The formation of black-hole analogues is examined for both short- and long-range interactions for arbitrary spatial dimensions greater than two. The former case includes non-linear derivative terms plus an inevitable external potential, while the latter one consists solely of gravity-like self-interactions for which intrinsic formation of analogue Schwarzschild-type background space-times is possible. The corresponding geometries are studied, and it is shown how they can be made stable. Their Hawking temperature is estimated, and it is found that in certain set-ups it can be significantly increased, thus providing better detectability.
\end{abstract}


\maketitle

\section{Introduction}
\label{sec:Introduction}

In recent years, Bose-Einstein condensates (BECs) have been subject to considerable study as well-suited analogue models of gravity \cite{Unruh:1981bi, Visser} (\cf~\cite{BLV} for an extended review and further references). It has been demonstrated that perturbations of the condensate are governed by the same equation of motion as scalar fields in curved space-time. In particular, transforming the associated effective metric into a metric conformally equivalent to the Schwarzschild metric offered the possibility of sonic black hole formation.

There is yet another field of application of BECs{\,---\,}namely on the level of a more complete microscopic picture of gravitational bound states \cite{Dvali}. These are, on the quantum level, described as macroscopic graviton condensates. Black holes being a prime example, wherefore the condensate description allows to successfully resolve the information paradox, to give a microscopic origin of Hawking radiation, and to explain the black hole's negative heat capacity \cite{Dvali}, just to mention a few. As the true spin-two (graviton) condensation involving all non-linear derivative interactions is rather difficult to handle, scalar toy models have been introduced, resembling certain aspects of black holes remarkably well \cite{Berkhahn, N-Portrait-related}. For correctly invoking the decay of black holes derivative interactions turn out to be essential.

One question to ask is now: How does the analogue gravity of generic scalar theories with derivative interactions look? Other questions arise: Does a horizon exist and where is it located? As BECs in the lab are always subject to an external potential: How does it have to look in order to generate analogue black-hole space-times? Furthermore, conceptually: Is it possible to form analogue black holes from intrinsic characteristics alone, possibly with a Hawking temperature that might be tuned to an easy detectable range? If yes, this may serve as an anchor model, on the one hand establishing a basis to easily observe black-hole quantum/semi-classical phenomena, as well as on the other hand providing a step towards a more complete microscopic picture of gravitational aspects. Furthermore, Boson stars (\cf~\eg~\cite{Boson-stars}) as well as the description of dark matter (halos) via BECs \cite{BEC-Dark-Matter, Rotating-BEC} provide two important astrophysical and even cosmological fields of application \cite{Kuhnel-Rampf}.

The structure of this work is as follows: In Sec.~\ref{sec:Set-Up} we introduce the theory by specifying the Hamiltonian for the Bose-Einstein condensate under consideration. Sec.~\ref{sec:Short--Range-Interactions} concerns the case of specific short-range self-interaction (involving quartic two- and four-derivative terms). The respective background and the fluctuations are studied, and the external potential is calculated. Furthermore, analytic expressions for the location of the respective horizon are derived. Then, in Sec.~\ref{sec:Long--Range-Interactions}, we introduce a gravity-like long-range interaction, for which a class of horizon geometries is derived in Subsec.~\ref{sec:Background-Long--Range}{\,---\,}without the need for any external potential. Stability of these solutions is studied in Subsec.~\ref{sec:Stability-Long-Range}. In Sec.~\ref{sec:Hawking-Temperature} we estimate the Hawking temperature and elaborate on its parameter dependence. Finally, we conclude in Sec.~\ref{sec:Summary-and-Outlook}.

\section{Set-Up}
\label{sec:Set-Up}

Our starting point is a $d$-dimensional Bose-Einstein condensate in an external potential, being described by the Hamiltonian
\begin{align}
\begin{split}
	\hat{\Hcal}
		&=
								\int\limits_{V}\!\d^{d}x\.
								\bigg\{
									\hat{\psi}^{\dagger}
									\Big(
										\Kcal
										+
										V_{\text{ext}}^{\mu}
									\Big)
									\hat{\psi}
									+
									\Ucal
									\big[
										\hat{\psi}
										,\.
										\ldots
									\big]
								\bigg\}
								\; ,
								\label{eq:hamiltonian}
\end{split}
\end{align}
with $\Kcal \defas - \frac{\hslash^{2}}{2m}\nabla^{2}$, and the subscript '$^{\mu}$' refers to the inclusion of a chemical potential $\mu$ into the external one, \ie, $V_{\text{ext}}^{\mu} \defas V_{\text{ext}} - \mu$. The condensate is supposed to be constituted by a number of $N \gg 1$ particles, and the (internal) potential $\Ucal$ shall, a priori, depend on $\hat{\psi}$, $\hat{\psi}^{\dagger}$, and spatial derivatives thereof. Furthermore, it may contain non-linearities as well as non-localities, as expected, \eg, from gravitational self-interaction. We work in units where the Planck constant $\hslash$ is equal to one, although we will keep its dependence explicitly in some expressions. Furthermore, energy is expressed in units of the chemical potential, and length shall be given in terms of the parameter $m$.

In Bose-Einstein condensates, the quantum state $\hat{\psi}$ consists of two components: a highly-occupied ground state, condensate part $\Psi \defas \langle | \hat{\psi} | \rangle$, which shall here be described by a classical field (due to the high occupation number), and a quantum-fluctuation part $\hat{\phi}$. In the so-called Madelung representation \cite{Madelung-Representation} the condensate part reads
\vs{-2mm}
\begin{align}
	\Psi
		&\equiv
								\sqrt{n_{0}}\,
								\ee^{\irm S}
								\; ,
\end{align}
with $n_{0}$ being the ground-state particle-number density, and the phase $S$ is connected to the velocity $v_{0}$ of the condensate through
\vs{-1mm}
\begin{align}
	v_{0}
		&=
								\frac{\hslash}{m}\.\nabla S
								\; .
\end{align}

\section{Short-Range Interactions}
\label{sec:Short--Range-Interactions}

As a first step we assume that below the critical temperature only interactions of zero range will contribute. Different form the original toy model used in Refs.~\cite{Dvali, N-Portrait-related} to study Bose-Einstein condensates of gravitons, we shall supplement with derivative interactions. The purpose of doing so is the inherent derivative nature of the gravitational self-interactions. In fact, in a rudimentary fashion, for $d = 1$, a quartic interaction of the form
\begin{align}
		&\sim
								\partial\hat{\psi}^{\dagger}
								\partial\hat{\psi}^{\dagger}
								\partial\hat{\psi}
								\partial\hat{\psi}
								\; ,
								\label{eq:U-short-range-pre}
\end{align}
has been included in the model of Ref.~\cite{Berkhahn} to mimic black-hole collapse, and in particular Hawking radiation, more realistically. Indications have been found that only when adding derivative interactions the black hole stays at the critical point of a quantum phase transition throughout the collapse.

However, the model of Ref.~\cite{Berkhahn} presents only a very first step in this direction, and the study of higher-dimensional models (in particular in $3 + 1$ dimensions) and also of more general interactions is in order. In this spirit, we will consider the $d$-dimensional interaction potential,
\vs{-1.5mm}
\begin{align}
	\Ucal
		\big[
			\hat{\psi}
			,\.
			\ldots
		\big]
		&=
								\frac{1}{2}\.U_{0}\;
								\hat{\psi}^{\dagger}\.\hat{\psi}^{\dagger}\.\hat{\psi}\.\hat{\psi}
								\notag
								\displaybreak[1]
								\\[0.5mm]
		&\hph{=}
								+
								\alpha\;\hat{\psi}^{\dagger}
								\big(
									\nabla\hat{\psi}^{\dagger}
								\big)
								\!\cdot\!
								\big(
									\nabla\hat{\psi}
								\big)\mspace{1mu}
								\hat{\psi}
								\notag
								\displaybreak[1]
								\\[1mm]
		&\hph{=}
								+
								\beta\.
								\big(
									\nabla\hat{\psi}^{\dagger}
								\big)
								\!\cdot\!
								\big(
									\nabla\hat{\psi}^{\dagger}
								\big)
								\big(
									\nabla\hat{\psi}
								\big)
								\!\cdot\!
								\big(
									\nabla\hat{\psi}
								\big)
								\notag
								\displaybreak[1]
								\\[1mm]
		&\hph{=}
								+
								\gamma\.
								\contraction{\mspace{11mu}}
								{
									1
								}
								{\big(
									\nabla\hat{\psi}^{\dagger}
								\big)
								\big(
									\nabla\hat{\psi}^{\dagger}
								\big)
								\mspace{13mu}}
								{\mspace{20mu}
								\big(
									\nabla\hat{\psi}
								\big)}
								\big(
									\nabla\hat{\psi}^{\dagger}
								\big)
								\big(
									\nabla\hat{\psi}^{\dagger}
								\big)
								\!\cdot\!
								\big(
									\nabla\hat{\psi}
								\big)
								\big(
									\nabla\hat{\psi}
								\big)
								\; ,
								\label{eq:U-short-range}
\end{align}
with real $U_{0}$, $\alpha$, $\beta$ and $\gamma$, and contractions are indicated in the standard fashion. 

Eq.~\eqref{eq:U-short-range} implies for the corresponding Heisenberg equation of motion,
\begin{align}
	\irm\partial_{t}\hat{\psi}
		&=
								\Big[
									\hat{\Hcal},\,
									\hat{\psi}
								\Big]
								\notag
								\\[1mm]
		&=
								\Big[
									\Kcal
									+
									V_{\text{ext}}^{\mu}
									+
									U_{0}|\hat{\psi}|^{2}
								\Big]
								\hat{\psi}
								\notag
								\\[1mm]
		&\mspace{23mu}
								-
								\alpha
								\Big[
									\hat{\psi}^{\dagger}
									\big(
										\triangle \hat{\psi}
									\big)\hat{\psi}
									+
									\hat{\psi}^{\dagger}
									\big(
										\nabla\hat{\psi}
									\big)^{2}
								\Big]
								\notag
								\\[1mm]
		&\mspace{23mu}
								-
								\beta\mspace{0.5mu}
								\contraction{\mspace{169mu}}{1}{\mspace{45mu}}{}
								\contraction{\mspace{243mu}}{1}{\mspace{37mu}}{}
								\Big[
									2\.
									\big(
										\triangle\hat{\psi}^{\dagger}
									\big)
									\big(
										\nabla\hat{\psi}
									\big)^{2}
									+
									4\.
									\big(
										\nabla\hat{\psi}^{\dagger}
									\big)
									\big[
										\nabla
										\big(
											\nabla\hat{\psi}
										\big)
									\big]
									\nabla\hat{\psi}
								\Big]
								\notag
								\\[1mm]
		&\mspace{23mu}
								-
								\gamma\mspace{0.5mu}
								\contraction{\mspace{30mu}}{1}{\mspace{77mu}}{}
								\contraction[1.8ex]{\mspace{53mu}}{1}{\mspace{98mu}}{}
								\Big[
									2\.
									\big[
										\nabla
										\big(
											\nabla\hat{\psi}^{\dagger}
										\big)
									\big]
									\big(
										\nabla\hat{\psi}
									\big)
									\big(
										\nabla\hat{\psi}
									\big)
								\notag
								\\[1mm]
		&\mspace{23mu}
								\phantom{-\gamma\Big[\;}
									+
									2\.
									\big(
										\nabla\hat{\psi}^{\dagger}
									\big)
									\!\cdot\!
									\big(
										\nabla\hat{\psi}
									\big)
									\big(
										\triangle\hat{\psi}
									\big)
								\notag
								\\[1mm]
		&\mspace{23mu}
								\phantom{-\gamma\Big[\;}
								\contraction{\mspace{44mu}}{1}{\mspace{45mu}}{}
								\contraction{\mspace{118mu}}{1}{\mspace{37mu}}{}
									+
									2\.
									\big(
										\nabla\hat{\psi}^{\dagger}
									\big)
									\big[
										\nabla
										\big(
											\nabla\hat{\psi}
										\big)
									\big]
									\nabla\hat{\psi}
								\Big]
								\. .
								\label{eq:eom}
								\\[1mm]
								\notag
\end{align}

\subsection{Background}
\label{sec:Background-Short--Range}

The (complex) condensate wave equation, Eq.~\eqref{eq:eom}, can be split into two real equations,
\begin{widetext}
\begin{subequations}
\begin{align}
		\partial_{t}n_{0}
		&=
								-
								\frac{ 1 }{ m }
								\left(
									\nabla S
								\right)
								\big(
									\nabla n_{0}
								\big)
								-
								\frac{1}{m}
								\left(
									\triangle S
								\right)
								n_{0}
								\notag
								\displaybreak[1]
								\\[1mm]
		&\hph{=}
								+							
								\Bigg[
									4
									n_{0}
									\big(
										\nabla n_{0}
									\big)
									\!\cdot\!
									\big(
										\nabla S
									\big)
									\Big[
										\beta
										\big(
											\nabla S
										\big)^{2}
										+
										\alpha
									\Big]
								\notag
								\displaybreak[1]
 								\\[1mm]
		&\hph{=}
								\hphantom{+\Bigg[\;\.}
									+
									2\.
									(
										\beta
										+
										\gamma
									)
									\big(
										\nabla n_{0}
									\big)
									\!\cdot\!
									\big(
										\nabla \nabla n_{0}
									\big)
									\!\cdot\!
									\big(
										\nabla S
									\big)
									+
									\gamma
									\big(
										\triangle S
									\big)
									\Big[
										\big(
											\nabla n_{0}
										\big)^{2}
										+
										4 n_{0}^{2}
										\big(
											\nabla S
										\big)^{2}
									\Big]
								\notag
								\displaybreak[1]
 								\\[2mm]
		&\hph{=}
								\hphantom{+\Bigg[\;\.}
 									+
									2 n_{0}^{2}
 									\Big[
										\big(
											\triangle S
										\big)
 										\Big(
											2 \beta
											\big(
												\nabla S
											\big)^{2}
											+
											\alpha
										\Big)
											+
										4
										(
											\beta
												+
											\gamma
										)
 										\big(
											\nabla S
										\big)
										\!\cdot\!
										\big(
											\nabla \nabla S
										\big)
										\!\cdot\!
 										\big(
											\nabla S
										\big)
									\Big]
								\displaybreak[1]
									\\[2mm]
		&\hph{=}
									\hphantom{+\Bigg[\;\.}
									+
									\big(
										\nabla n_{0}
									\big)^{2}
 									\Big[
										\beta\.
										n_{0}^{-1}
 										\big(
											\nabla n_{0}
										\big)
										\!\cdot\!
 										\big(
											\nabla S
										\big)
										-
										\beta
										\big(
											\triangle S
										\big)
									\Big]
 								\notag
								\displaybreak[1]
									\\[2mm]
 		&\hph{=}
									\hphantom{+\Bigg[\;\.}
 									+
									\big(
										\nabla n_{0}
									\big)
									\!\cdot\!
									\bigg(
										2 \beta
										\Big[
 											\big(
												\nabla \nabla S
											\big)
											\!\cdot\!
											\big(
												\nabla n_{0}
											\big)
 											-
											\big(
												\nabla \nabla n_{0}
											\big)
											\!\cdot\!
 											\big(
												\nabla S
											\big)
										\Big]
								\notag
								\displaybreak[1]
 									\\[0.5mm]
		&\hph{=}
									\hphantom{+\Bigg[\;\.+\big(\nabla n_{0}\big)\bigg(\;\.}
										+
										\big(
											\nabla S
										\big)
										\Big[
											2
											\beta
											\big(
												\triangle n_{0}
											\big)
											+
											4
											(
												\beta
												+
												2 \gamma
											)
											n_{0}
											\big(
											\nabla S
												\big)^{2}
 											-
											n_{0}^{-1}
											\beta
											\big(
												\nabla n_{0}
											\big)^{2}
										\Big]
 									\bigg)
								\Bigg]
 								\label{eq:real1}
								\displaybreak[1]
								\\[3mm]
\intertext{and}
	- \partial_{t}S
		&=
								\frac{ 1 }{ \sqrt{n_{0}\.} }\.\Kcal \sqrt{n_{0}\.}
								+
								\frac{1}{2m}
								\left(
									\nabla S
								\right)^{2}
								+
								V_{\text{ext}}^{\mu}
								+
								U_{0}\.n_{0}
								\notag
								\displaybreak[1]
								\\[1mm]
		&\hph{=}
								-
								\frac{1}{2 n_{0}^{2}}
								\Bigg[
									(
										\beta
										+
										\gamma
									)
									n_{0}
									\big(
										\nabla n_{0}
									\big)^{2}
									\big(
										\nabla S
									\big)^{2}
									-
									\frac{ 1 }{ 2 }
									\gamma
									\big(
										\triangle n_{0}
									\big)
									\big(
										\nabla n_{0}
									\big)^{2}
									+
									4 \alpha
									n_{0}^3
									\big(
										\nabla S
									\big)^{2}
									+
									4
									(
										\beta
										+
										\gamma
									)
									n_{0}^{3}
									\big(
										\nabla S
									\big)^{4}
								\displaybreak[1]
								\notag
								\\[1mm]
	&\hph{=}
 								\mspace{70mu}
										-
										n_{0}^{2}
										\bigg(
											\big(
												\triangle n_{0}
											\big)
											\Big[
												\alpha
												-
												2 \beta
												\big(
													\nabla S
												\big)^{2}
											\Big]
											+
											2
											\gamma
											\big(
												\triangle n_{0}
											\big)
											\big(
												\nabla S
											\big)^{2}
								\displaybreak[1]
								\notag
 									\\[1mm]
	&\hph{=}
 									\mspace{73mu}
									\hphantom{-n_{0}^{2}\Big(\;\.}
											+
											4 \beta
											\big(
												\nabla S
											\big)
											\!\cdot\!
											\Big[
												\big(
													\nabla \nabla n_{0}
												\big)
												\!\cdot\!
												\big(
													\nabla S
												\big)
												-
												\big(
													\nabla \nabla S
												\big)
												\!\cdot\!
												\big(
													\nabla n_{0}
												\big)
											\Big]
										\bigg)
								\displaybreak[1]
 								\label{eq:real2}
								\\[3mm]
 		&\hph{=}
 								\mspace{70mu}
 								-
								\frac{ 1 }{ 4 n_{0} }
								\big(
									\nabla n_{0}
								\big)^{2}
								\bigg(
									2 n_{0}
									\Big[
										\beta
										\big(
											\triangle n_{0}
										\big)
										+
										2
										(
											\beta
											+
											\gamma
										)
										n_{0}
										\big(
											\nabla S
										\big)^{2}
									\Big]
									-
									3
									(
										\beta
										+
										\gamma
									)
									\big(
										\nabla n_{0}
									\big)^{2}
								\bigg)
								\displaybreak[1]
								\notag
 								\\[3mm]
 	&\hph{=}
 								\mspace{70mu}
 								-
 								\big(
									\nabla n_{0}
								\big)
								\!\cdot\!
								\bigg(
									(
										\beta
 										+
										\gamma
									)
 									\big(
										\nabla \nabla n_{0}
									\big)
									\!\cdot\!
									\big(
										\nabla n_{0}
									\big)
									+
									4
									n_{0}^{2}
									\Big[
										(
											\beta
 											+
											\gamma
										)
 										\big(
											\nabla \nabla S
										\big)
										\!\cdot\!
 										\big(
											\nabla S
										\big)
										+
										\beta
											\big(
											\nabla S
										\big)
										\big(
 											\triangle S
										\big)
									\Big]
									\bigg)
								\Bigg]
								.
								\notag
\end{align}
\end{subequations}
\end{widetext}

As a check, we note that for $\alpha = \beta = \gamma = 0$ the first equation reduces to the standard equation for the current conservation,
\vs{-3mm}
\begin{subequations}
\begin{align}
	\partial_{t}n_{0}
	+
	\nabla\!\cdot\!\jbm
		&=
								0
								\; ,\mspace{-59.5mu}
								\label{eq:current-conservation-equation}
\end{align}
where $\jbm \defas n_{0}\.v_{0}$, while the second equation can be identified with Bernoulli's equation,
\begin{align}
	\partial_{t}S
	+
	\frac{ 1 }{ \sqrt{n_{0}\.} }\.\Kcal \sqrt{n_{0}\.}
	+
	V_{\text{ext}}^{\mu}
	+
	U_{0}\.n_{0}
	+
	\frac{ m }{ 2 }\.v_{0}^{2}
		&=
								0
								\; .
								\label{eq:Bernoullis-equation}
\end{align}
\end{subequations}
The second term on the left-hand side is of purely quan-\\tum-mechanical origin.\\

\subsection{Fluctuations}
\label{sec:Fluctuations-Short--Range}

After having briefly discussed the background level, we next study the behavior of the fluctuation $\hat{\phi}$ at first order. Omitting the 'hats' and setting
\begin{subequations}
\begin{align}
	\rho
		&\defas
								\sqrt{n_{0}\.}
								\big(
									\ee^{-\irm S}\phi
									+
									\ee^{\irm S}\phi^{\dagger}
								\big)
								\; ,
								\\[2mm]
	\Phi
		&\defas
								\frac{\hslash}
								{2\.m\.\irm\sqrt{n_{0}\.}}
								\big(
									\ee^{-\irm S}\phi
									-
									\ee^{\irm S}\phi^{\dagger}
								\big)
								\; ,
\end{align}
\end{subequations}
we form two equations (adding on subtracting Eq.~\eqref{eq:eom} and its daggered pendent) and express $\phi$ and $\phi^{\dagger}$ in terms of $\rho$ and $\Phi$. Then, applying the hydrodynamical approximation (\cf~Ref.~\cite{BLV}), we use one of the resulting equations to solve for $\rho$ (in dependence of $\Phi$). Plugging this into the other equation, we obtain a closed dynamical equation for $\Phi$ alone. We will discuss aspects of it in the following subsections.

\subsubsection{Without Derivative Interactions}
\label{sec:alpha-=-beta-=-0}

Let us first consider the case $\alpha = \beta = \gamma = 0$, which corresponds to the usual Bose-Einstein condensate. Then, under the assumption that the spatial variation of $\rho $ is small, we find
\begin{align}
	\big(
		\partial_{t}
		+
		\nabla v_{0}
	\big)
	\frac{m}{U_{0}}
	\big(
		\partial_{t}
		+
		v_{0}\nabla
	\big)
	\Phi
		&\simeq
								\nabla
								\big(
									n_{0}
									\nabla
								\big)
								\Phi
								\; .
\end{align}
Now, following Ref.~\cite{Kurita:2008fb}, introducing a symmetric matrix,
\begin{align}
	\big(
		g_{\mu\nu}
	\big)
		&\propto
								\left(
									\begin{array}{ccccc}
										-
										\left(
											\frac{U_{0}n_{0}}{m}
											-
											v_{0}^{2}
										\right)						& - v_{0}^{1}	& \cdots
																	& \cdots		& - v_{0}^{d}\\
										- v_{0}^{1}						& 1			& 0
																	& \cdots		& 0\\
										\vdots						& 0			& \ddots
																	& \ddots		& \vdots\\
										\vdots						& \vdots		& \ddots
																	& \ddots		& 0\\
										- v_{0}^{d}						& 0			& \cdots
																	& 0			& 1
									\end{array}
								\right)
								,
								\label{eq:metric}
\end{align}
the above equation of motion for the phase fluctuations can be rewritten as
\begin{align}
	\frac{1}{\sqrt{-\det{g}\.}}\.\partial_{\mu}
	\Big[
		\sqrt{-\det{g}\.}\.g^{\mu\nu}\partial_{\nu}\Phi
	\Big]
		&\simeq
								0
								\; .
								\label{eq:Klein-Gordon-Equation}
\end{align}
This equation is formally and kinematically analogous to the Klein-Gordon equation for massless scalar fields in fixed curved $d$-dimensional space-time if $g_{\mu\nu}$ is identified with a space-time metric \footnote{One should, however, keep in mind that here this analogy is purely formal as the matrix $g_{\mu \nu}$ does not transform appropriately under general coordinate transformations.} via ${\d}s^{2} = g_{\mu\nu}{\d}x^{\mu}{\d}x^{\nu}$, where $\mu$, $\nu = 0, 1, \ldots, d$.

We note that if $v_{0} m / U_{0} n_{0} = r_{\text{S}} / r$ (where $r \defas | x |$ and $r_{\text{S}} / r > 0$), Eq.~\eqref{eq:metric} describes a metric which is conformally equivalent to the Schwarzschild metric in so-called Painlev{\'e}-Gullstrand (PG) coordinates \cite{PG}. In three spatial dimensions, these are related to the usual Schwarzschild (S) coordinates via
\begin{align}
	t_{\text{PG}}
		&=
								t_{\text{S}}
								-
								4 M\.\mathrm{arctanh}
								\Bigg[
									\sqrt{\frac{ 2\.M }{ r }\,}\.
								\Bigg]\mspace{-3mu}
								+
								2\.\sqrt{2 M r\.}
								\; ,
\end{align}
where $M$ is the black-hole mass, being connected to the Schwarzschild horizon $r_{\text{S}}$ via $M = c^{2}\.r_{\text{S}} / 2\.G$, with $G$ being Newton's constant, and $c$ is the speed of light.

Taking $n_{0}$ to be constant, and applying the static and isotropic limit, Eq.~\eqref{eq:current-conservation-equation} yields
\begin{align}
	v_{0}( r )
		&=
								\frac{v_{00}}{r^{d - 1}}
								\; .
								\label{eq:v0-n0-relation}
\end{align}
Here the constant $v_{00}$ is determined, \eg, by fixing the initial flow of the condensate at a certain distance from the origin. Then, given the effective metric Eq.~\eqref{eq:metric}, a horizon (at $r = r_{*}$) occurs if
\begin{align}
	g_{00}( r )\Big|_{r = r_{*}}
		&\propto
								\bigg[
									\frac{ U_{0}\.n_{0} }{ m }
									-
									v_{0}( r )^{2}
								\bigg]
								\Bigg|_{r = r_{*}}
		\overset{!}{=}				0
								\; .
\end{align}
This implies
\begin{align}
	r_{*}
		&=
								\bar{M}^{\frac{ 1 }{ 2 d - 2 }}
								\; ,
								\label{eq:rstar}
\end{align}
where
\vs{-2mm}
\begin{align}
	\bar{M}
		&=
								\frac{ m\.v_{00}^{2} }
								{ n_{00}\.U_{0}}
								\; .
								\label{eq:Mbar}
\end{align}
Eq.~\eqref{eq:real2} allows us to obtain an expression for the external potential,
\vs{-1mm}
\begin{align}
	V_{\text{ext}}^{\mu}( r )
		&=
								-
								\frac{ 1 }{ 2 }\.m\.v_{00}^{2}\.
								\frac{ 1 }{ r^{2 d - 2} }
								-
								n_{00}\.U_{0}
								\; .
								\label{eq:Vextmu}
\end{align}
Like the metric, it has a rather simple structure.

As expected, Eq.~\eqref{eq:rstar} shows that the black hole becomes larger if the strength of the attractive self-interaction is lowered. We will continue to observe this physically meaningful behaviour also in the other cases.

\subsubsection{With Derivative Interactions}
\label{sec:alpha-ne-beta-ne-0}

Let us now come to the case with derivative interactions being present. Therefore, we again apply the static isotropic limit, and furthermore consider the case of constant background density. Then, Eq.~\eqref{eq:real1} yields
\begin{align}
	\triangle S
		&=
								-
								\frac{
									8\.m\.n_{0}
									(
									\beta
										+
									\gamma
									)
 									\big(
										\nabla S
									\big)
									\!\cdot\!
									\big(
										\nabla \nabla S
									\big)
									\!\cdot\!
 									\big(
										\nabla S
									\big)
								}{
									1
									+
									2\.m\.n_{0}\.\alpha
									+
									4\.m\.n_{0}
									(
										\beta
											+
										\gamma
									)
									(
										\nabla S
									)^{2}
								}
								\; .
								\label{eq:Laplace-S}
\end{align}
This suggests to take
\vs{-1.5mm}
\begin{align}
	\gamma
		&=
								-
								\beta
								\; ,
								\label{eq:gamma=-beta}
\end{align}
for which $\triangle S = 0$. We will adopt this choice throughout the reminder of this work. Then, as before, we have
\begin{align}
		&
								\notag
								\\
	v_{0}( r )
		&=
								\frac{ v_{00} }{ r^{d - 1} }
								\; ,
								\label{eq:v0(r)}
								\\
								\notag
\end{align}
with some constant $v_{00}$. This implies, with Eq.~\eqref{eq:real2},
\begin{align}
\begin{split}
								V_{\text{ext}}^{\mu}( r )
		&=
								-
								\frac{ m\.v_{00}^{2} }{ 2 }
								\frac{ (1 + 4\.\alpha\.m\.n_{0}) }{ r^{2d - 2} }
								-
								U_{0} n_{0}
								.
								\label{eq:Vextmu-alpha-beta}
\end{split}
\end{align}

Before coming to the general case, it is instructive to set $\beta = 0$ first. There we find for the purely temporal component of the metric
\begin{align}
	g_{00}( r )
		&\propto
								1
								-
								\frac{ v_{00}^{2} }{ r^{2d - 2} }
								\bigg[
									\frac{ n_{0}\.U_{0} }{ m }\.
									(
										1
										+
										2\.m\.n_{0}\.\alpha
									)
								\bigg]^{-1}
								\times
								\notag
								\displaybreak[1]
								\\[2mm]
		&\hph{\propto}
								\notag
								\\
		&\hph{\propto}
								\phantom{1-\;\.}
								\times
								\Big[
									1
									+
									6\.m\.n_{0}\.\alpha\.
									(
										1
										+
										2\.m\.n_{0}\.\alpha
									)
								\Big]
								\; ,
								\label{eq:g00-alpha}
\end{align}
which implies that the horizon is located at $r = r_{*}$, with
\begin{align}
	r_{*}
		&=
								\sqrt[2 d - 2]
								{
									\frac{
										m\.v_{00}^{2}
										+
										6\.\alpha\.m\.n_{0}
										(
											1
											+
											2\.m\.n_{0}\.\alpha
										)\.
										m\.v_{00}^{2}
										}
										{
										n_{0}\.U_{0}
										(
											1
											+
											2\.m\.n_{0}\.\alpha
										)}
								\,}
								\; .
								\label{eq:rstar-alpha-beta}
\end{align}
Hence, for large positive $\alpha$ the horizon grows. It can easily be checked that all of the above equations contain their respective counterparts of the previous subsection.

\begin{figure}[t]
	\centering
	\includegraphics[scale=0.42]{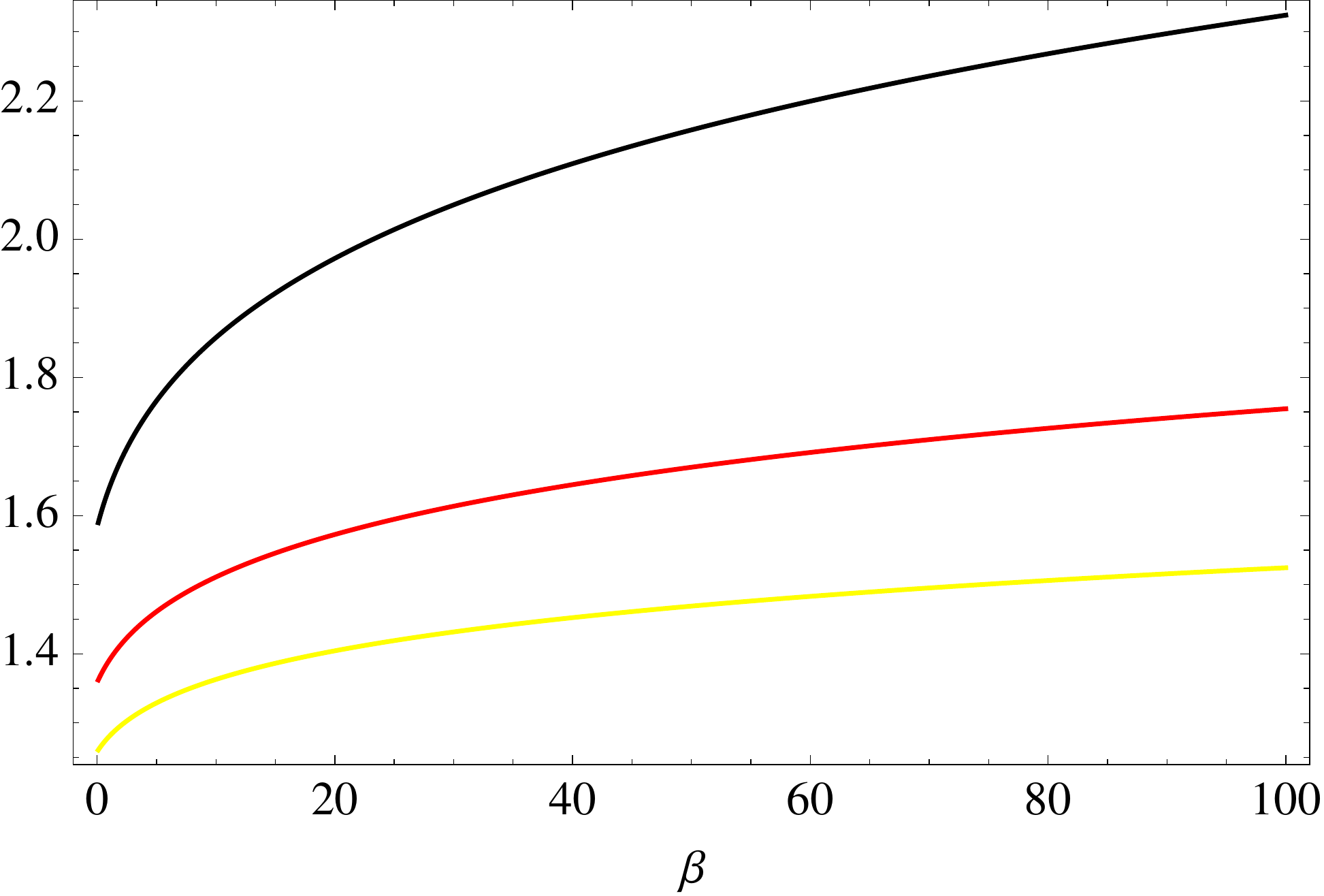}
	\caption{$r_{*}$ as a function of $\beta$
			for three different spatial dimensions
			$d$ ($3$, $4$, $5$; bottom to top).
			The other parameters are
			$\alpha = m = n_{0} = | v_{00} | = U_{0} = 1$.}
	\label{fig:rstar-alpha-beta}
\end{figure}

For also $\beta \ne 0$ the situation becomes more involved. One finds that the horizon in this general case is located at a radius $r = r_{*}$, with
\begin{widetext}
\begin{align}
	r_{*}
		&=
								\frac{ 8^{\frac{1}{2 d - 2}} }{
									\beta m^4\.n_{0} v_{00}^4
									(
										4\.\alpha\.m\.n_{0}
										+
										1
									)
								}
								\Bigg[
									\.\sqrt{m^2\.v_{00}^4
										\Big(
											16\.\beta\.m^2\.n_{0}^2
											U_{0}
											(
												2 \alpha\.m\.n_{0}
												+
												1
											)
											(
												4\.\alpha\.m\.n_{0}
												+
												1
											)
											+
											\big[
												6\.\alpha\.m\.n_{0}
												(
													2\.\alpha\.m\.n_{0}
													+
													1
												)
												+
												1
											\big]^2
										\Big)
									\,}
								\displaybreak[1]
								\notag
 								\\[3mm]
		&\hph{=}
								\mspace{198mu}
									-
									m\.v_{00}^2
									\big[
										6\.\alpha\.m\.n_{0}
										(
											2\.\alpha\.m\.n_{0}
											+
											1
										)
										+
										1
									\big]
 								\Bigg]^{\frac{1}{2 - 2 d}}
								\; .
								\label{eq:rstar-alpha-beta}
\end{align}
\end{widetext}
Fig.~\ref{fig:rstar-alpha-beta} depicts this solution for $r_{*}$ as a function of $\beta$ for three different spatial dimensions.

\section{Long-Range Interactions}
\label{sec:Long--Range-Interactions}

The main focus of the previous study of general derivative interaction was on investigating more refined models for the set-up of Bose-Einstein condensates of gravitons, proposed Dvali and Gomes \cite{Dvali}. As derivative interaction are essential to understand the black-hole collapse, we studied a respective $d$-dimensional model, generalizing the simple short-range derivative set-up of Ref.~\cite{Berkhahn}.

Going now from short- to long-range interactions, the focus shall here be on two other fields of application: One is that of Boson stars (\cf~\eg~\cite{Boson-stars}), and the other one consists of Bose-Einstein condensate dark-matter halos \cite{BEC-Dark-Matter}. In the latter the condensate consists of a (ultra-light) particle, which can, for instance, be axion-like. Interesting new and potentially constraining phenomena and effects are expected, as for instance lensing is different, and, if the halo rotates, vortices might form (\cf, \eg, \cite{Rotating-BEC, Rotating-pure-BEC}).

Most models for BEC dark-matter halos consist of a (complex) scalar field with or without local self-coupling, but with a gravity-induced self-interaction. This is, however, not only of astro-physical importance. In fact, certain configurations of intense off-resonant laser beams can effectively induce an attractive $1 / r$ potential between the condensate atoms \cite{self-bound}. Such interactions can mimic gravitational-like interactions and lead to stable self-bound systems with unique properties.

Furthermore, as far as analogue gravity is concerned, arguments from the previous section suggest that it is impossible to form a horizon in a Bose-Einstein condensate without an external potential if only local interactions of the general form \eqref{eq:U-short-range} are considered. This statement seems likely to hold for all short-range interactions. However, it might be that an external potential is not compulsory in the case of long-range interactions. This shall be one aspect of what we are going to investigate next.

In this spirit we are particularly interested in such $1 / r^{d - 2}$-interactions,
\vs{-1.5mm}
\begin{align}
	\Ucal
		\big[
			\hat{\psi}^{\dagger}
			,\.
			\hat{\psi}
		\.\big]
		&=
								\frac{ 1 }{ 2 }
								\int \!\d^{d}y\;
								\hat{\psi}^{\dagger}( x )\.\hat{\psi}^{\dagger}( y )
								\frac{C}{| y - x |^{d - 2}}_{}\.
								\hat{\psi}( y )\.\hat{\psi}( x )
								\displaybreak[1]
								\notag
								\\[1mm]
		&\hph{=}
								-\,
								\mu\.
								\hat{\psi}^{\dagger}( x )
								\hat{\psi}( x )
								\; ,
								\label{eq:Hbec}
								\\[-7mm]
								\notag
\end{align}
with some (positive) coupling constant $C$.

As it is gravity in the case of BEC dark-matter halos which determines $C$, in the lab, it depends on various quantities, for instance the wavenumber $q$ and the intensity $I$ of the laser, the polarizability $\alpha$ of the atoms as well as on their number $N$. For example in the case of Natrium, on obtains for $C \approx 10^{-13}\.{\rm eV}\.{\rm nm}$ for $d = 3$, $q \approx 1 / \mu \mrm$, and $I \approx 10^{8}\.\Wrm / \mrm^{2}$ \cite{self-bound}.

At first sight it seems that for astro-physical BECs a gravity-induced interaction potential of the form \eqref{eq:Hbec} cannot give rise to stable Bose-Einstein condensates. However, as shown \eg~in Refs.~\cite{SN} the gravitational-like interaction may not necessarily induce a collapse of the condensate. This can, for instance, be seen from the scaling of the kinetic energy versus that of the potential.

Now, our theory \eqref{eq:Hbec} can be rephrased as
\begin{subequations}
\begin{align}
	\irm\.\partial_{t} \psi
		&=
								\!
								\bigg[
									-
									\frac{1}{2\.m}\.\triangle
									+
									\Phi\.
								\bigg]
								\psi
								\; ,
								\label{eq:eom-2}
\end{align}
with
\vs{-2mm}
\begin{align}
	\triangle \Phi
		=&
								\.-
								\Omega_{d}\.
								\rho
								\; ,
								\label{eq:Delta-Phi}
								\\[2mm]
	\rho
		\defas&
								\;\.
								\psi^{\dagger}\psi
								\; ,
								\label{eq:rho}
\end{align}
\end{subequations}
where we used
\begin{align}
	\triangle\.\frac{ 1 }{ | x - y |^{d - 2} }
		&=
								-\,\Omega_{d}\.\delta^{d}( x - y )
								\; ,
								\label{eq:Laplace-on-1/r**d-1/2}
\end{align}
with $\Omega_{d} \defas 2\.\pi^{d} / \Gamma[ d / 2 ]$, $\Gamma[\,\cdot\,]$ being the gamma function, and set $\mu = 0$.

Hence the system (\ref{eq:eom-2}-c) describes a wave-function in its own (self-sourced) gravitational potential. Therefore it constitues a so-called Schr{\"o}dinger-Newton system \cite{Karolyhazy:1966zz}. The mechanism for these systems to admit stable solution is the counter-balancing of the spread of the wave function by the attractive self-interaction.

\subsection{Background}
\label{sec:Background-Long--Range}

As above, in the static case, the Gross-Pitaevskii equation, implied by Eq.~\eqref{eq:Hbec}, yields
\begin{align}
	\nabla\!
	\cdot\!
	\big[
		n_{0}( r )\.v_{0}( r )
	\big]
		&=
								0						
								\; ,
								\label{eq:real1-LR}
\end{align}
which leads to
\vs{-1mm}
\begin{align}
	v_{0}( r )
		&=
								\frac{v_{00}}{r^{d - 1}}\.\frac{ 1 }{ n_{0}( r ) }
								\; ,
								\label{eq:v0-n0-relation-LR}
\end{align}
and allows to obtain an integral equation for the particle-number density alone,
\begin{align}
	\mu
		&\simeq
								\frac{ 1 }{ 2 }\.m\mspace{1mu}v_{00}^{2}\.\frac{ 1 }{ r^{2 d - 2}}\.\frac{ 1 }{ n_{0}^{2}( r ) }
								+
								C\! \int \!\d^{d} y\.
								\frac{n_{0}( | y | )}{| y - x |^{d - 2}}
								\; .
								\label{eq:real2-LR-2}
\end{align}
Using relation \eqref{eq:Laplace-on-1/r**d-1/2} we find
\begin{align}
	\bar{n}_{0}( r )
		&\simeq
								\frac{ 1 }{ r^{d - 1} }\.\partial_{r}\.r^{d - 1}\.\partial_{r}\mspace{-1mu}
								\Bigg[
									\frac{1}{r^{2 d - 2}}\.\frac{ 1 }{ \bar{n}_{0}^{2}( r ) }
								\Bigg]
								\. ,
								\label{eq:real2-LR-diff-1}
\end{align}
where we defined
\vs{-2mm}
\begin{subequations}
\begin{align}
	\bar{n}_{0}( r )
		&\defas
								n_{0}( r )\,
								\bar{c}^{\,1 / 3}
								\; ,
								\label{eq:n0(r)}
\intertext{and set}
	\bar{c}
		&\defas
								\frac{ 1 }{ 2 }\.\bar{m}\.v_{00}^{2}\.
								\; ,
								\label{eq:bar-c}
\end{align}
\end{subequations}
with $\bar{m} \defas m / ( \Omega_{d}\.C )$.

\begin{figure}[t]
	\centering
	\includegraphics[scale=0.42]{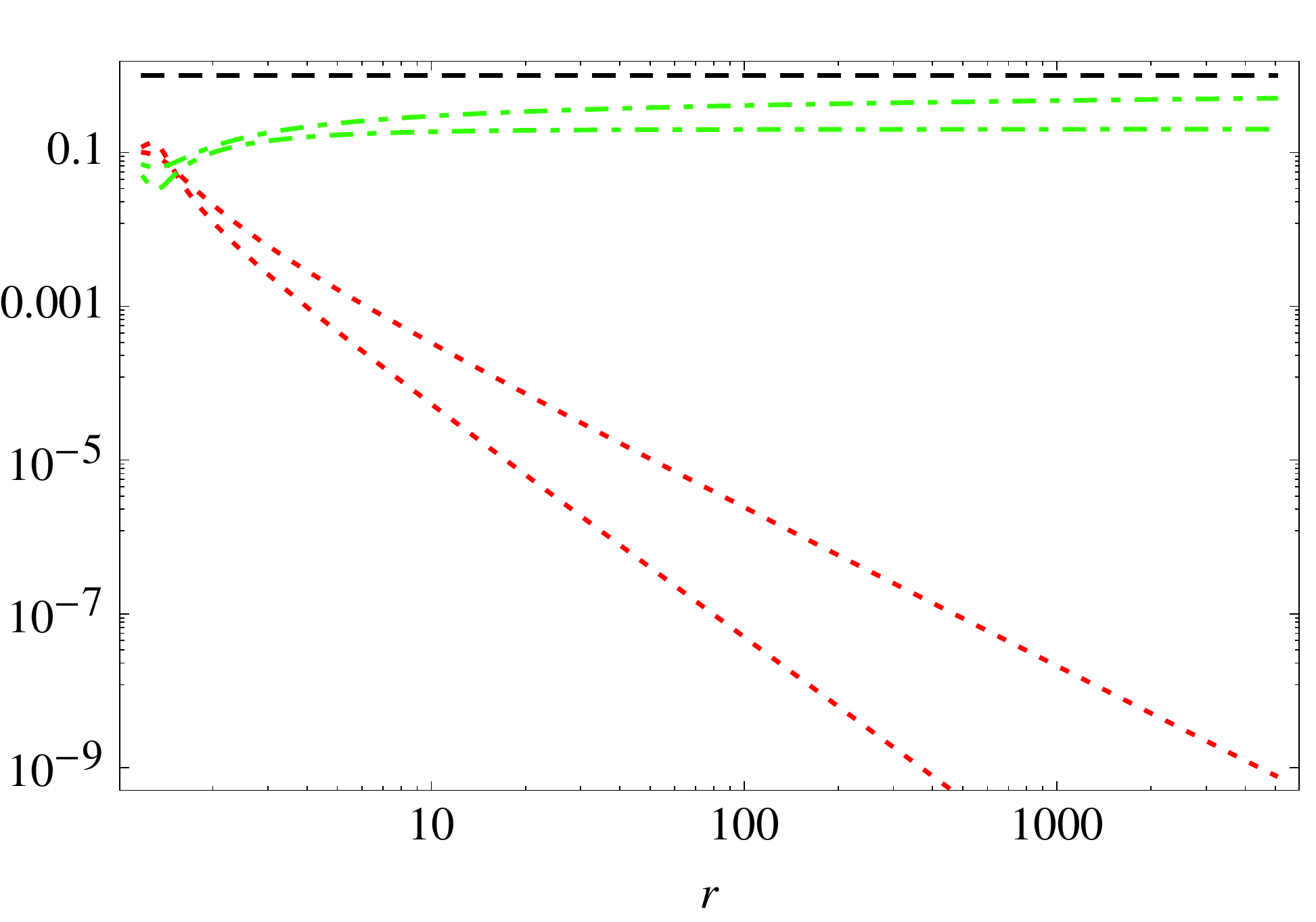}
	\caption{Particle-number density $n_{0}( r )$ ({\it red, dashed$\mspace{1mu}$})
			and velocity $| v_{0}( r ) |$ ({\it green, dot-dashed\.})
			for $d = 3$ ({\it upper respective graphs$\mspace{1mu}$})
			and $d = 4$ ({\it lower respective graphs$\mspace{1mu}$})
			for $r_{\text{max}} = 5 \cdot 10^{4}$, $m = 10^{-1}$, $| v_{00} | = 10^{-2}$, and $C = 10^{-4}$.}
	\label{fig:n0-v0-LR-d-dependence}
	\vs{2mm}
\end{figure}

\begin{figure}[t]
	\centering
	\includegraphics[scale=0.42]{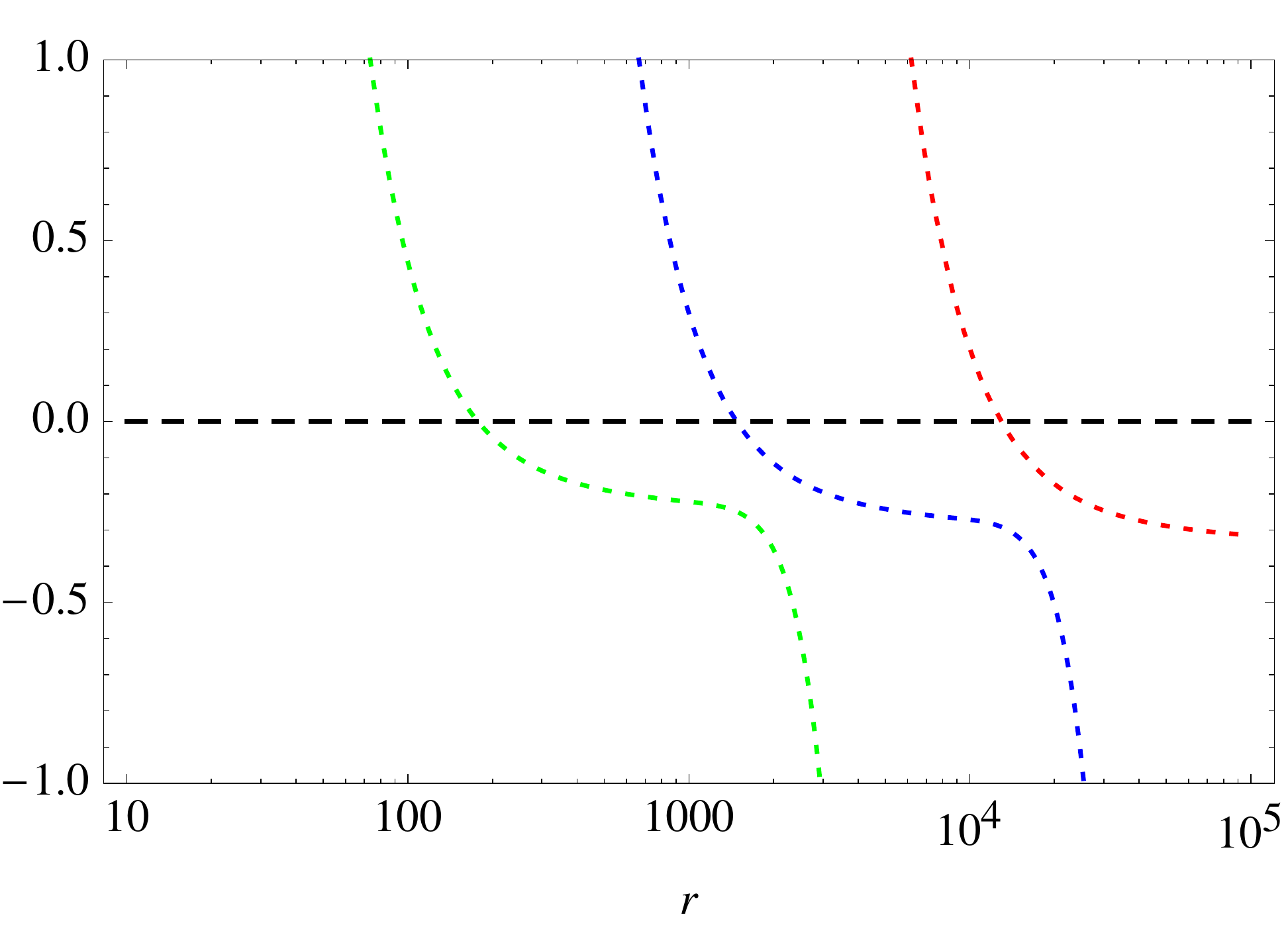}
	\caption{Metric component $g_{00}( r )$
			for various values of $r_{\text{max}}$
			($10^{3}$, $10^{4}$, $10^{5}$; left to right).
			The other parameters are $d = 3$, $m = 10^{-1}$, $| v_{00} | = 10^{-2}$, and $C = 10^{-4}$.}
	\label{fig:g00-LR-d=3-rmax-dependence}
\end{figure}

Fig.~\ref{fig:n0-v0-LR-d-dependence} shows particular solutions to Eq.~\eqref{eq:real2-LR-diff-1}, plus the corresponding velocity $v_{0}( r )$ [which is related to $n_{0}( r )$ via Eq.~\eqref{eq:v0-n0-relation-LR}] for the case of $d = 3$ and $d = 4$, respectively. The situation is the following: The particles clump about the centre of the black hole, and their number density falls off away from its origin. At the same time the velocity grows till it saturates at a constant level, implying that the aforementioned fall-off in $n_{0}( r )$ goes like $r^{1 - d}$.

From Fig.~\ref{fig:g00-LR-d=3-rmax-dependence} we see how the metric component $g_{00}( r )$ behaves as a function of radius, and how it scales with the condensate size. For the given parameters, it is located at about a tenth of its extent, and the larger it is, the better is the hydro-dynamical approximation.

Further numerical results are that smaller values of $| v_{00} |$ correspond to smaller chemical potential. The same holds true for the coupling $C$, and also for variation of $m${\,---\,}facts that will be utilized in Sec.~\ref{sec:Hawking-Temperature}, when we estimate the Hawking temperature. Evaluating the \rhs~of Eq.~\eqref{eq:real2-LR-2}, using the numerical solution for $n_{0}( r )$ from before, we checked that $\mu$ appears to be really constant.

It is also clear that increasing the coupling $C$ makes $n_{0}( r )$ more peaked at the centre of the analogue black hole, which is expected form the attractivity of the self-interaction. Furthermore, it can be checked that all of the above findings are essentially insensitive to any set of specific, allowed and reasonable initial conditions.

Summarized, for the long-range model \eqref{eq:Hbec}, we have shown that it is indeed possible to establish an analogue space-time describing a black hole \textit{solely from intrinsic quantities}. On top of that, this statement holds true for arbitrary dimensions $d > 2$ (\cf~Fig.~\ref{fig:n0-v0-LR-d-dependence}).\\[-7mm]

\subsection{Stability}
\label{sec:Stability-Long-Range}

Having derived the background solutions, it is tempting to perform a stability analysis. Expanding the particle density-number as well as the phase up to second order, \footnote{Note that we allow the perturbations $\delta n$ and $\delta S$ to have\\ a general coordinate dependence, and not just assume spherical symmetry, as we did for the background. This is important, because the system might be stable under perturbations that respect its symmetry, and unstable with regard to those that do not.}
\begin{subequations}
\begin{align}
	n( x , t )
		&=
								n_{0}( r )
								+
								\delta n( x , t )
								+
								\Ocal\mspace{-1mu}
								\big(
									\delta n^{2}
								\big)
								\; ,
								\label{eq:nexpand}
								\\[2mm]
	S( x , t )
		&=
								S_{0}( r )
								+
								\delta S( x , t )
								+
								\Ocal\mspace{-1mu}
								\big(
									\delta S^{2}
								\big)
								\; ,
								\label{eq:sexpand}
\end{align}
\end{subequations}
the first-order equations of motion (using the zeroth-order condensate equations) read
\begin{subequations}
\begin{align}
	\partial_{t}\delta n( x , t )
		&\simeq
								-\.
								\nabla\!\cdot\!
								\bigg(
									v_{0}( r )\,
									\delta n( x , t )
									+
									\frac{ 1 }{ m }\.
									n_{0}( r )\,
									\nabla\delta S( x , t )
								\bigg)
								,
								\label{eq:deltan}
								\displaybreak[1]
								\\[2mm]
\begin{split}
	\partial_{t}\delta S( x , t )
		&\simeq
								-\,
								v_{0}( r )
								\cdot
								\nabla \delta S( x , t )
								\\[2mm]
		&\hph{\simeq}
								-
								\int \!\d^{d}y\.
								\frac{C}{| y - x |^{d - 2}}\.
								\delta n( y , t )
								\. .
								\label{eq:deltas-1}
\end{split}
\end{align}
\end{subequations}
Acting with $\triangle$ on Eq.~\eqref{eq:deltas-1} and employing formula \eqref{eq:Laplace-on-1/r**d-1/2}, we find
\vs{-1mm}
\begin{align}
	\delta n( x , t )
		&\simeq
								\frac{ 1 }{ \Omega_{d}\.C }\,
								\triangle
								\bigg[
									\partial_{t}\delta S( x , t )
									+
									v_{0}( r )
									\cdot
									\nabla \delta S( x , t )
								\bigg]
								\. .
								\label{eq:delta-n-in-terms-of-delta-S}
\end{align}
Plugging Eq.~\eqref{eq:delta-n-in-terms-of-delta-S} into Eq.~\eqref{eq:deltan} yields an equation solely for $\delta S$,
\vs{-1mm}
\begin{align}
		&					
								\triangle
								\bigg[
									\partial^{2}_{t}\delta S( x , t )
									+
									v_{0}( r )
									\cdot
									\nabla \partial_{t} \delta S( x , t )
								\bigg]
								\notag
								\displaybreak[1]
								\\
		&\qq\simeq
								-\.
								\nabla\!\cdot\!
								\bigg(
									v_{0}( r )\;
									\triangle
									\bigg[
										\partial_{t}\delta S( x , t )
										+
										v_{0}( r )
										\cdot
										\nabla \delta S( x , t )
									\bigg]
								\notag
								\displaybreak[1]
								\\
		&
								\hphantom{\qq=-\.\nabla\!\cdot\!\bigg(\.}
									+
									\frac{ \Omega_{d}\.C }{ m }\.
									n_{0}( r )\,
									\nabla\delta S( x , t )
								\bigg)
								\. ,
								\vphantom{\Bigg|}
								\label{eq:only-delta-S}
\end{align}
which is clearly a higher-derivative equation. Now, since, for large $r$, the velocity $v_{0}( r )$ approaches a constant, \ie, $n_{0}( r )$ becomes proportional to $1 / r^{d - 1}$ [\cf~Eq.~\eqref{eq:v0-n0-relation-LR}], we are approximately left with
\vs{-1mm}
\begin{align}
	\Big[
		\partial_{t}
		+
		v_{0}( r )\!\cdot\!\nabla
	\Big]^{2}\.
	\triangle\delta S( x, t )
		&\simeq
								0
								\; .
								\label{eq:Xi-Eq-large-r}
\end{align}
The quantity $\triangle\delta S( x, t )$ can be thought of as the gradient of the velocity perturbation. Hence, any pulse keeps its shape at large distance from the centre of the black hole, and if it is large, the above behavior also includes regions well within the horizon. Along similar lines, Eq.~\eqref{eq:deltan} becomes
\vs{-2mm}
\begin{align}
	\Big[
		\partial_{t}
		+
		v_{0}( r )\!\cdot\!\nabla
	\Big]\.
	\delta n( x, t )
		&\simeq
								0
								\; .
								\label{eq:Xi-Eq-large-r}
\end{align}
Thus, our stability result also holds for arbitrary pertur-bations in the particle-number density. Moreover, these findings are true for all dimensions greater than two.

\section{Hawking Temperature}
\label{sec:Hawking-Temperature}

Having studied analogue black-hole-like objects above, it is tempting to estimate their specific Hawking temperature. Contrary to certain folklore, Hawking radiation neither requires the validity of the Einstein equations nor does it probe an underlying theory of quantum gravity. \cite{Visser:2001kq} The essential requirement, namely a horizon geometry on which quantized perturbations evolve, are perfectly met in the set-up under consideration.

Setting $d = 3$, and following the pioneering work of Unruh \cite{Unruh:1981bi}, the Hawking temperature $T_{\text{H}}$ behaves approximately as
\begin{align}
	T_{\text{H}}
		&\approx
								\frac{ 1 }{ 2 \pi k_{\text{B}} }\.
								\partial_{r}\big| v_{0}( r ) \big| \bigg|_{r\,=\,r_{*}}
								\; ,
								\label{eq:T-Hawking-1}
\end{align}
where $k_{\text{B}}$ is the Boltzmann constant, and the horizon is located at $r = r_{*}$.

Of course, for a more precise estimate, it would be necessary to accurately and fully determine the fluctuations' dispersion relation, which is modified by the presence of the long-range interactions as compared to the standard short-range case (\cf~\cite{Sakar}). This has some effect on the spectrum of the Hawking-temperature (\cf~\cite{Corley:1996ar}). We understand that Eq.~\eqref{eq:T-Hawking-1} gives a first rough estimate.

From the behavior of $v_{0}( r )$ (\cf~Fig.~\ref{fig:n0-v0-LR-d-dependence}), we observe that the bigger the analogue black hole, the lower $T_{\text{H}}$, which is expected. It is possible to derive an approximate analytical formula for the analogue Hawking temperature,
\vs{-2mm}
\begin{align}
	T_{\text{H}}
		&\sim
								10^{-11}
								\left(
								\frac{ v_{00} }{ r_{*} }
									\big[
										\srm^{-1}
									\big]
								\right)\!
								\left(
									\frac{ C }{ \hslash\.v_{00} }
								\right)^{\!1 / 3}\.
								\Krm
								\; .
								\label{eq:T-Hawking-2}
\end{align}
Hence, the value of the Hawking temperature depends on the velocity $v_{00}$, the radius $r_{*}$ (which is about a tenth of the condensate extent $r_{\text{max}}$), and the coupling $C$.

It is now tempting to evaluate $T_{\text{H}}$ for concrete physical systems. Let us start with Bose-Einstein condensate dark matter halos. Taking exemplary $\vphantom{1^{1}_{_{1}}}v_{00} \approx 100\.{\rm km} / \srm$, $r_{*} \approx 100\.{\rm kpc}$, and $C = m^{2}\.G_{\text{N}}$, where $G_{\text{N}} \approx 10^{-10}\.\mrm^{3}\.\srm^{-2}\.\krm\grm^{-1}$ is Newton's constant, we find
\begin{align}
	T_{\text{H}} \approx \big( m [\krm\grm] \big)^{2 / 3} \.10^{-19}\.\Krm
								\; ,
								\label{eq:T-Hawking-BEC-dark-matter-halo}
\end{align}
which seems to be far beyond present (and near future) detectability, even for heavy bosons, and a small and fast rotating halo. This originates essentially from two aspects: the typically rather large $r_{*}$ and the smallness of the gravitational coupling.\footnote{The former point is improved for Boson stars (going from galactic scales to essentially solar size and below); Also, the rotation is typically much faster. However, the resulting analogue Hawking temperature will still remain tiny.}


It is precisely these points where laboratory systems can anchor. Taking, again exemplary, the value of the coupling mentioned earlier, $C \approx 10^{-13}\.{\rm eV}\.{\rm nm}$ \cite{self-bound}, as well as $v_{00} \approx 1\.\mrm / \srm$ and $r_{*} \approx 10^{-6}\.\mrm$, one finds
\begin{align}
	T_{\text{H}} \approx 10^{-7}\.\Krm
								\; .
								\label{eq:eq:T-Hawking-BEC-lab}
\end{align}
This is still small, but much larger than the analogue Hawking temperature of $T_{\text{H}} \approx 10^{-12}\.\Krm$, which has been experimentally inferred for a different (and even much noisier) set-up consisting of specific water waves \cite{Weinfurtner:2013cd}.

In the Bose-Einstein condensate set-up under consideration, the coupling $C$ depends, in particular, on the intensity $I$ of the off-resonant laser beams which induce the attractive $1 / r$ potential. Here, $C$ is approximately proportional to $I$, which offers a way to increase the analogue Hawking temperature further.

\section{Summary \& Outlook}
\label{sec:Summary-and-Outlook}

In the first part of this work we studied general non-relativistic Bose-Einstein condensates with short-range derivative interactions. These are relevant as possible toy models for the recently-proposed graviton-condensation picture of space-time \cite{Dvali}. We investigated the encoded analogue geometry, which involves the calculation of the horizon for any dimension greater than two in dependence on the model parameters. Furthermore, we derived an expression for the external potential which is necessary to achieve these analogue geometries, thus providing a basis to prepare and to test aspects of those new graviton BEC scenarios in laboratory.

In the second part, we have extended our analysis also to long-range interactions, which are relevant on the one hand to Boson stars and Bose-Einstein condensate dark matter halos, and, on the other hand, to laboratory systems with laser-induced long-range interaction. Here we also found that analogue black-hole horizons form. These are classically stable if parameters are such that they are large. This approximate stability is in full agreement with the standard intuition.

We have estimated the analogue Hawking temperature which is associated with the horizon, and found that on astrophysical scale Hawking radiation exists but with an extremely low temperature. This situation changes drastically when laboratory systems are considered, mainly because the self-coupling is not given by the tiny gravitational interaction, but rather by a tunable coupling which is many orders of magnitude larger. It would be very interesting to investigate this aspect further.

Furthermore, in the case of rotating BECs, the effect of so-called super-radiance \cite{Super-Radiance}, \ie, the amplification of certain scattered waves through the extraction of rotational energy, should occur. In a forthcoming publication we will elaborate more on this interesting aspect \cite{Kuhnel-Rampf}, which might help to observationally constrain BEC dark-matter halo models.

Finally, from a conceptual viewpoint, we have demonstrated one possibility to achieve{\,---\,}without any external potential, \ie, just from system-inherent properties alone{\,---\,}a connection between non-relativistic Bose-Einstein condensates with long-range interactions and a relativistic Schwarzschild-like analogue space-time, in arbitrary spatial dimensions $d > 2$.\\



\acknowledgements

{\color{white}.}\\[-7mm]
It is a pleasure to thank Felix Berkhahn, Gia Dvali, Cristiano Germani, Ariel Goobar, Stefan Hofmann, Florian Niedermann, Cornelius Rampf, Robert Schneider, and in particular Sophia M{\"u}ller for fruitful discussion and support. I would like to thank the anonymous Referee for valuable remarks. This work was in parts supported by the DFG cluster of excellence ’Origin and Structure of the Universe’, the Alexander von Humboldt Foundation, and the Swedish Research Council (VR) through the Oskar Klein Centre. FK acknowledges hospitality of the Arnold Sommerfeld Center in Munich, where parts of this work was done.


\end{document}